\documentclass[fleqn,usenatbib]{mnras}

\usepackage[T1]{fontenc}
\usepackage{ae,aecompl}

\usepackage{graphicx}	% Including figure files
\usepackage{amsmath}	% Advanced maths commands
\usepackage{color}

\usepackage{mathptmx}
\usepackage{txfonts}

\newcommand{\msolar}{\;{\rm M}_{\odot}}

\newcommand{\swift}{{\sc Swift}}
\newcommand{\eagle}{{EAGLE}}
\newcommand{\gadget}{{\sc Gadget}}

\title[Sub-grid models and P-SPH]{Inconsistencies arising from the coupling of galaxy formation sub-grid models to Pressure-Smoothed Particle Hydrodynamics}
\author[Borrow et al.]{
Josh Borrow$^{1}$,
Matthieu Schaller$^{2}$, and
Richard G. Bower$^{1}$
\\$^1$ Institute for Computational Cosmology, Department of Physics, University of Durham, South Road, Durham, DH1 3LE, UK
\\$^2$ Leiden Observatory, Leiden University, PO Box 9513, NL-2300 RA Leiden, The Netherlands
}

\begin{document}

\maketitle

\begin{abstract}Smoothed Particle Hydrodynamics (SPH) is a Lagrangian method for solving the
fluid equations that is commonplace in astrophysics, prized for its natural
adaptivity and stability. The choice of variable to smooth in SPH has been the
topic of contention, with smoothed pressure (P-SPH) being introduced to reduce
errors at contact discontinuities relative to smoothed density schemes.
Smoothed pressure schemes produce excellent results in isolated hydrodynamics
tests; in more complex situations however, especially when coupling to the
`sub-grid' physics and multiple time-stepping used in many state-of-the-art
astrophysics simulations, these schemes produce large force errors that can
easily evade detection as they do not manifest as energy non-conservation.
Here two scenarios are evaluated: the injection of energy into the
fluid (common for stellar feedback) and radiative cooling. In the former
scenario, force and energy conservation errors manifest (of the same order as
the injected energy), and in the latter large force errors that change rapidly
over a few timesteps lead to instability in the fluid (of the same order as the
energy lost to cooling). Potential ways to remedy these issues are explored
with solutions generally leading to large increases in computational cost.
Schemes using a Density-based formulation do not create these instabilities and
as such it is recommended that they are preferred over Pressure-based solutions
when combined with an energy diffusion term to reduce errors at contact
discontinuities.

\end{abstract}


\section{Introduction}

Over the past three decades, the inclusion of hydrodynamics in (cosmological)
galaxy formation simulations has become commonplace \citep{Hernquist1989,
Evrard1994, Springel2002, Springel2005, Dolag2009}. One of the first
hydrodynamics methods to be used in such simulations was Smoothed Particle
Hydrodynamics \citep[SPH, ][]{Gingold1977, Monaghan1992}. SPH is prized for
its adaptivity, conservation properties, and stability and is still used in
state-of-the-art simulations by many groups today \citep{Schaye2015,
Teklu2015, McCarthy2017, Tremmel2017, Cui2019, Steinwandel2020}; see
\citet{Vogelsberger2020} for a recent overview of cosmological simulations.

As the SPH method has developed, two key issues have arisen. The first, a
consequence of the non-diffusive nature of the SPH equations, was that the
method was unable to capture shocks. This was resolved by the addition of a
diffusive `artificial viscosity' term \citep{Monaghan1983}. This added
diffusivity is only required in shocks, and so many schemes include
particle-carried switches for the viscosity \citep{Morris1997,Cullen2010} to
prevent unnecessary conversion between kinetic and thermal energy in e.g.
shearing flows. The second, artificial surface tension appearing in contact
discontinuities \citep[e.g.][]{Agertz2007}, has led to the development of
several mitigation procedures. One possible solution is artificial
conductivity (also known as energy diffusion) to smooth out the discontinuity
\citep[e.g.][]{Price2008, Read2012, Rosswog2019}; this method applies an
extra equation of motion to the thermodynamic variable to transfer energy
between particles. The alternative solution, generally favoured in the
cosmology community, is to reconstruct a smooth pressure field
\citep{Ritchie2001, Saitoh2013, Hopkins2013}. This smooth pressure field
allows for a gradual transition pressure between hot and cold fluids,
suppressing any variation in the thermodynamic variable at scales smaller
than the resolution limit. This can be beneficial in fluids where there is a
high degree of mixing between phases, such as in gas flowing into galactic
haloes \citep[e.g.][]{ Tumlinson2017,Stern2019}.

Cosmological simulations typically include so-called `sub-grid' physics that
aims to represent underlying physics that is below the (mass) resolution
limit \citep[which is usually around $10^{3-7}
\msolar{}$;][]{Vogelsberger2014, Schaye2015, Hopkins2018, Marinacci2019,
Dave2019}. This is commonplace in many fields, and is essential in galaxy
formation to reproduce many of the observed properties of galaxies. One key
piece of sub-grid physics is star formation, which occurs on mass scales
smaller than a solar mass. Cold, dense, gas is required to enable stars to
form; to reach these temperatures and densities radiative cooling (which
occurs on atomic scales) must be included in a sub-grid fashion. Finally,
when these stars have reached the end of their life some will produce
supernovae explosions, which are modelled using sub-grid `feedback' schemes
\citep[such a sub-grid scheme is chosen for many reasons, including but not
limited to limited resolution and the 'overcooling problem'; see ][and
references for more information]{Navarro1993, DallaVecchia2012}. Each of
these processes has an impact on the hydrodynamics solver which must be
carefully examined. Here we employ a simple galaxy formation model including
implicit cooling and energetic feedback, based on the \eagle{} galaxy
formation model \citep{Schaye2015}, to understand how the inclusion of such a
model may affect simulations employing Density- or Pressure-based SPH
differently. We note, however, that the results obtained in the following
sections are applicable to all kinds of galaxy formation models, including
those that instead use instantaneous or `operator-split' cooling.

The rest of this paper is organised as follows: In \S \ref{sec:sph} the SPH
method is described, along with the Density- and Pressure-based schemes; in
\S \ref{sec:eagle} the basics of a galaxy formation model are discussed in
more detail; in \S \ref{sec:energyinjection} issues relating to injection of
energy into Pressure-based schemes are explored; in \S \ref{sec:eom} the SPH
equations of motion are discussed; in \S \ref{sec:timeintegration} the
time-integration schemes used in cosmological simulations are presented and
issues with sub-grid cooling are explored, and in \S \ref{sec:conclusions} it
is concluded that while Pressure-SPH schemes can introduce significant errors
it is possible in some cases to use measures (albeit computationally
expensive ones) to remedy them. Because of this added expense it is suggested
that a Density-based scheme is preferred, with an energy diffusion term used
to mediate contact discontinuities.
\section{Smoothed Particle Hydrodynamics}
\label{sec:sph}

\begin{figure*}
    \centering
    \includegraphics{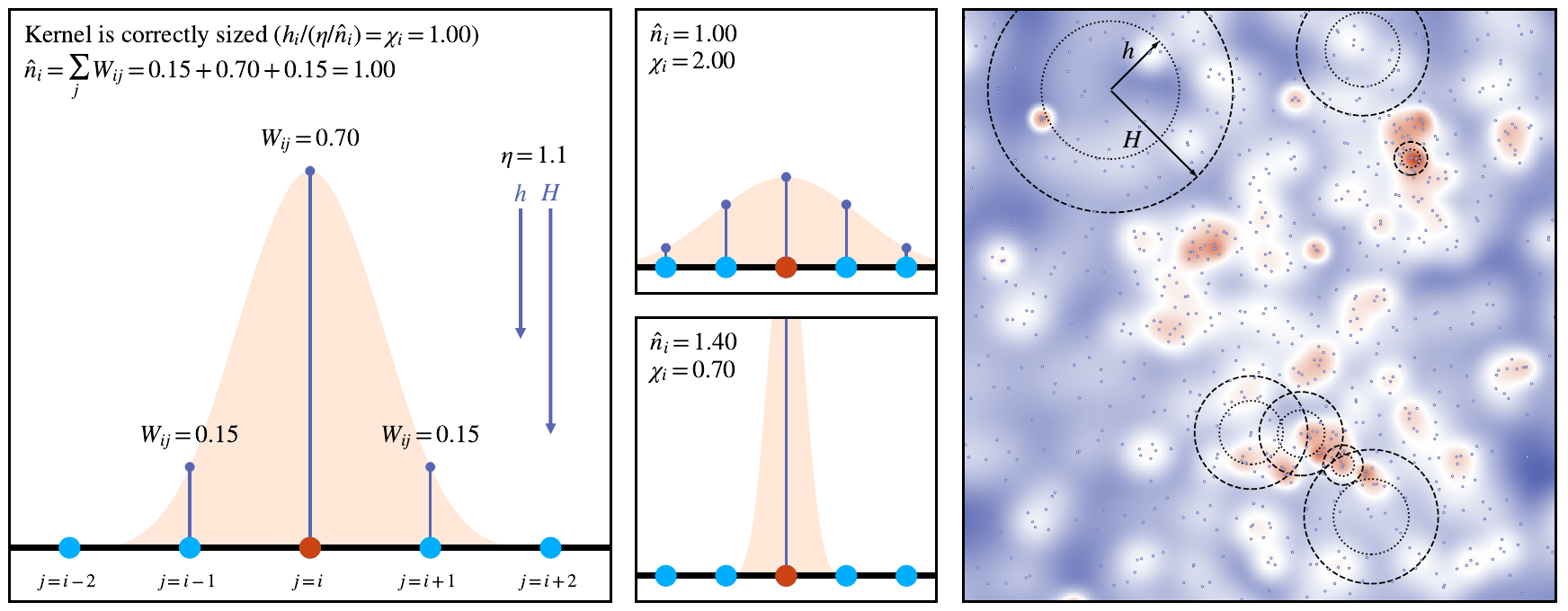}
    \caption{The three leftmost panels show the consequences of choosing a
    correct (large, left), too large (top right), and too small (bottom
    right) smoothing length (for $\eta = 1.1$) in 1D on a set of particles
    with an expected density $\hat{n} = 1$. This is quantified through both
    the density, $\hat{n}$, for the central particle $i$, and the ratio
    between the chosen smoothing length $h_i$ and the expected smoothing
    length given by $\eta / \hat{n}_i$, parametrised as $\chi_i$. $\chi_i$ is
    a well behaved function of the smoothing length, and finding the root of
    $\chi_i - 1$ is a reliable way to choose the value of $h_i$ that
    corresponds to a given choice of $\eta$. Note how the density is only
    erroneous in the case with a smoothing length that is too small (bottom
    panel); the larger smoothing length (top panel) produces the correct
    density but would be less computationally efficient and inconsistent with
    the chosen value of $\eta$. The rightmost panel shows a 2D case with a
    random particle distribution, with the background colour map showing the
    low (blue) to high (white and then red) density regions and the
    associated variation in smoothing length. Here, for selected particles,
    the smoothing length $h$ and kernel cut-off radius $H$ are shown with
    dotted and dashed lines respectively. In particular, note how the higher
    density regions show smaller smoothing lengths such that Equation
    \ref{eqn:numberdensity} is respected.
    }
    
    \label{fig:sph_description}
\end{figure*}

\begin{table*}
    \centering
    \begin{tabular}{l|c|c|l}
        Parameter Name & Symbol & Symbolic Units & Description \\
        \hline
        Number of dimensions & $n_D$ & None & Number of spatial dimensions (1-3) \\
        Particle position & $\mathbf{r}_i$ & [l] & Cartesian vector position of particles \\
        Inter-particle separation & $r_{ij} = |\mathbf{r}_i - \mathbf{r}_j|$ & [l] & Euclidean distance between two particles $i$ and $j$ \\
        Smoothing length & $h_i$ & [l] & Particle-carried smoothing length corresponding to FWHM of gaussian \\
        Number density & $\hat{n}_i$ & [l]$^{-n_D}$ & Local particle number density, i.e. the local density of particles \\
        Kernel & $W_{ij} = W(r_{ij}, h_i)$ & [l]$^{-n_D}$ & Value of the kernel at the distance between particles $i$ and $j$ \\
        Eta & $\eta$ & None & Ratio of local inter-particle separation to smoothing length \\
        Kernel support ratio & $\gamma_K$ & None & Ratio between cut-off radius of the kernel and smoothing length; kernel-dependent \\
        Ratio of specific heats & $\gamma$ & None & The ratio of specific heats for the ideal gas, here $\gamma = C_P / C_V = 5/3$ \\
        Cut-off radius & $H_i = \gamma_K h_i$ & [l] & Maximal radius at which the kernel takes a non-zero value \\
        Density & $\hat{\rho}_i$ & [m][l]$^{-n_D}$ & Local mass density, here defined at the position of particle $i$ \\
        Particle mass & $m_i$ & [m] & The mass of a particle, here the mass of particle $i$ \\
        Internal energy & $u_i$ & [l]$^{2}$[t]$^{-2}$ & Internal energy \emph{per unit mass} of particle $i$ \\
        Entropy & $A_i$ & [m]$^{1-\gamma}$[l]$^{3\gamma - 1}$[t]$^{-2}$ & Entropy \emph{per unit mass} of particle $i$ \\
        Pressure & $P_i$ or $\hat{P}_i$ & [m][l]$^{-1}$[t]$^{-2}$ & Pressure of the field, either at particle positions (left) or smoothed (right) \\
        Sound-speed & $c_{{\rm s}, i}$ & [l][t]$^{-1}$ & Speed of sound at the position of particle $i$ \\
        Weighted density & $\bar{\rho}_i$ & [m][l]$^{-3}$ & Smoothed pressure-weighted density, i.e. $\hat{P}_i / ((\gamma - 1) u_i)$ \\
        Particle velocity & $\mathbf{v}_i$ & [l][t]$^{-1}$ & Cartesian vector velocity of the particle $i$ \\
        {\it h}-factor & $f_{ij}$ & None & Correction factor for variable smoothing lengths (between particles $i$ and $j$) \\
    \end{tabular}
    \caption{Table of symbols used in the rest of the paper. Symbols are
    defined, along with their units (in terms of unit mass [m], unit length
    [l], and unit time [t]) here.}
    \label{tab:symbols}
\end{table*}

SPH is a Lagrangian method that uses particles to discretise the fluid. To
find the equation of motion for the system, and hence integrate a fluid in
time, the forces acting on each particle are required. In a fluid, these
forces are determined by the local pressure field acting on the particles.
The ultimate goal of the SPH method, then, is to find the pressure gradient
associated with a set of discretised particles; once this is obtained finding
the equations of motion is a relatively simple task. The reader is referred
to the first few pages of the review by \citet{Price2012} for more information
on the fundamentals of the SPH method.

Before continuing, it is important to separate the two types of quantities
present in SPH. The first, \emph{particle carried properties} (denoted as
symbols with an index corresponding to their particle, e.g. $m_i$ is the mass
of particle $i$), are valid only at the positions of particles in the system
and include variables such as mass. The second, \emph{field properties}
(denoted as symbols with a hat, and with a corresponding index if they are
evaluated at particle positions, such as $\hat{\rho}_i$, the density at the
position of particle $i$), are valid at all points in the computational
domain, and generally are volumetric quantities. These field properties are
built out of particle-carried properties by convolving them with the
smoothing kernel.

The smoothing kernel is a weighting function of two parameters,
inter-particle separation ($|\mathbf{r}_i - \mathbf{r}_j| = r_{ij}$) and
smoothing length $h_i$, with a shape similar to a Gaussian with a full-width
half maximum of $\sqrt{2\ln 2} h_i$. The smoothing lengths of particles are
chosen such that, for each particle, the following equation is satisfied:
\begin{equation}
    \hat{n}_i = \sum_{{\rm All ~ particles}~ j} W(r_{ij}, h_i) = \left(
      \frac{\eta}{h_i}\right)^{n_D},
    \label{eqn:numberdensity}
\end{equation}
where $\hat{n}_i$ is the local number density, $n_D$ the number of spatial
dimensions, and the kernel $W(r_{ij}, h_i)$ (henceforth written as $W_{ij}$)
has the same dimensions as number density, typically being composed of a
dimensionless weighting function $w_{ij} = w(r_{ij} / h_i)$ such that $W_{ij}
\propto w_{ij} h_i^{-n_D}$. $\eta$ is a dimensionless parameter that
determines how smooth the field reconstruction should be (effectively setting
the spatial resolution), with larger values leading to kernels that encompass
more particles and typically takes values around $\eta \approx
1.2$\footnote{This corresponds to the popular choice of around 48 neighbours
for a cubic spline kernel.}. An important distinction is the difference
between the smoothing length, $h_i$, related to the full-width half-maximum
(FWHM) of the Gaussian that the kernel approximates, and the kernel cut-off
radius $H_i$. This cut-off radius is parametrised as $H_i = \gamma_K h_i$,
with $\gamma_K$ a kernel-dependent quantity taking values around $1.5-2.5$,
such that $H_i$ gives the maximum value of $r_{ij}$ at which the kernel will
be non-zero\footnote{The choice of which variable to store, $h$ or $H$, is
tricky; $h$ is more easily motivated \citep{Dehnen2012} and independent of the
choice of kernel, but $H$ is much more practical in the code as outside this
radius interactions do not need to be considered.}. We note that Table
\ref{tab:symbols} shows all symbols used regularly throughout this paper and
encourage readers to refer to it when necessary.

An example kernel \citep[the cubic spline kernel, see][for significantly more
information on kernels]{Dehnen2012} is shown in Fig.
\ref{fig:sph_description}, with three choices for the smoothing length that
satisfy Equation \ref{eqn:numberdensity}: one that is too large; one that is `just
right' for the given choice of $\eta$, and one that is too small. The choice to
satisfy both equations is not strictly equivalent to ensuring that the kernel
encompasses a fixed number of neighbouring particles; note how the edges of the
kernel in the left panel do not coincide with a particle, even despite their
uniform spacing.

To evaluate the mass density of the system, at the particle positions, the
kernel is again used to re-evaluate the above equation now including the
particle masses such that the density
\begin{equation}
    \hat{\rho}_i = \sum_{j} m_j W_{ij}
    \label{eqn:sphdensity}
\end{equation}
is the sum over the kernel contributions and neighbouring masses $m_j$ that
may differ between particles. Note that this summation includes the
self-contribution from the particle $i$, $m_i W(0, h_i)$.

Typically in SPH, the particle-carried property of either internal energy
$u_i$, or entropy $A_i$ (per unit mass)\footnote{Note that this quantity is not
really the `entropy', but rather the adiabat that corresponds to this choice
of entropy, hence the choice of symbol $A$.} is chosen to encode the thermal
properties of the particle. These are related to each other, and the
particle-carried pressure, through the ideal gas equation of state
\begin{equation}
    P_i = (\gamma - 1)u_i\hat{\rho}_i = A_i\hat{\rho}^{\gamma},
    \label{eqn:equationofstate}
\end{equation}
with the ratio of specific heats $\gamma = C_P / C_V = 5/3$ for the fluids
usually considered in cosmological hydrodynamics models.

Alternatively, it is possible to construct a smooth pressure field that
is evaluated at the particle positions such that
\begin{equation}
    \hat{P}_i = \sum_j (\gamma - 1) m_j u_j W_{ij} = \left(\sum_j m_j A_j^{1/\gamma} W_{ij}\right)^\gamma,
    \label{eqn:smoothedpressure}
\end{equation}
directly includes the particle-carried thermal quantities of the
neighbours into the definition of the pressure.

The differences between SPH models that use the particle pressures evaluated
through the equation of state and smoothed density (i.e. those that use
Equations \ref{eqn:sphdensity} and \ref{eqn:equationofstate}), known as
Density SPH, and those that use the smooth pressures (i.e. those that
use Equation \ref{eqn:smoothedpressure}), known as Pressure SPH, is the central
topic of this paper. Frequently, the SPH scheme is also referred to
by its choice of thermodynamic variable, internal energy or entropy,
as Density-Energy (Density-Entropy) or Pressure-Energy (Pressure-Entropy).

SPH schemes are usually implemented as a fixed number of `loops over
neighbours' (often just called loops). For a basic scheme like the ones
presented above, two loops are usually used. The first loop, frequently called
the `density' loop, goes over all neighbours $j$ of all particles $i$ to
calculate their SPH density (Equation \ref{eqn:sphdensity}) or
smooth pressure (Equation \ref{eqn:smoothedpressure}). The second loop,
often called the `force' loop, evaluates the equation of motion for
each particle $i$ through the use of the pre-calculated smoothed
quantities of all neighbours $j$. Each loop is computationally expensive,
and so schemes that require extra loops are generally unfavourable
unless they provide a significant benefit. State-of-the-art schemes typically
use three loops, inserting a `gradient' loop between the `density'
and `force' loops to calculate either improved gradient estimators
\citep{Rosswog2019} or coefficients for artificial viscosity and
diffusion schemes \citep{Price2008, Cullen2010}.

\section{A simple galaxy formation model}
\label{sec:eagle}

The discussion that follows requires an understanding of two pieces of a
galaxy formation model: energy injection into the fluid and energy removal
from the fluid. These are used to model the processes of supernovae and AGN
feedback, and radiative cooling respectively. The results presented here are
not necessarily tied to the model used, and are applicable to a wide range of
current galaxy formation models that use Pressure-based SPH schemes. Here we
use a simplified version of the \eagle{} galaxy formation model as an
instructive example, as this used Pressure-Entropy SPH for its hydrodynamics
model in \citet{Schaye2015} and associated works \citep[of particular note is
][that discusses the effects of the choice of numerical SPH scheme on galaxy
properties]{Schaller2015}.

\subsection{Cooling}
\label{sec:cooling}

%Radiative cooling routines are provided as part of the \eagle{} model
%that use the \citet{Wiersma2009} tables to compute cooling rates based
%on particle density, temperature, and metallicity. These tables are
%interpolated implicitly, such that over a given time-step changes in
%the cooling rate may be incorporated.

The following equation is solved implicitly for each particle
separately:
\begin{equation}
    u(t + \Delta t) = u(t) + \frac{\mathrm{d}u}{\mathrm{d}t}(t)\Delta t,
\end{equation}
where $\mathrm{d}u/\mathrm{d}t$ being the `cooling rate' calculated from
the underlying atomic processes with the resulting final internal energy being
transformed into an average rate of change of internal energy as a function
of time over the step,
\begin{equation}
    \bar{\frac{\mathrm{d}u}{\mathrm{d}t}} = \frac{u(t + \Delta t) - u(t_i)}{\Delta t}.
    \label{eqn:avg_cooling_rate}
\end{equation}
After this occurs, this rate is limited in some circumstances
\citep[see][for more detail]{Schaye2015} that are not relevant to the
discussion here. This average `cooling rate' is then applied as either
an addition to the $\mathrm{d}u/\mathrm{d}t$ or $\mathrm{d}A/\mathrm{d}t$
from the hydrodynamics scheme for each particle depending on the variable
that the scheme tracks.

\subsection{Energy Injection Feedback}

A common, simple, feedback model is implemented as heating particles \emph{by} a
constant temperature jump. It is possible to implement different types of feedback
with this method, all being represented with a separate change in temperature
$\Delta T$. For supernovae feedback, $\Delta T_{\rm SNII} = 10^{7.5}$ K, and
for AGN $\Delta T_{\rm AGN} = 10^{8.5}$ K (in \eagle{}). The change in
temperature does not actually ensure that the particle has this temperature
once the feedback has taken place, however; the amount of energy
corresponding to heating a particle from 0 K to this temperature is added to
the particle. This ensures that even in cases where the particle is hotter
than the heating temperature energy is still injected.

To apply feedback to a given particle, this change in temperature must
be converted to a change in internal energy. This is performed by using
a linear relationship between temperature and internal energy to find the internal
energy that corresponds to a temperature of $\Delta T$, and adding this
additional energy onto the internal energy of the particle.

\section{Energy injection in Pressure-Entropy}
\label{sec:energyinjection}

In cosmology codes it is typical to use the particle-carried entropy as the
thermodynamic variable rather than the internal energy. This custom originated
because in many codes \citep[of particular note here is \gadget{};][]{
Springel2005} the choice of co-ordinates in a space co-moving with
expansion due to dark energy is such that the entropy variable is
cosmology-less, i.e. it is the same in physical and co-moving space. Entropy is
also conserved under adiabatic expansion, meaning that fewer equations of
motion are required. This makes it convenient from an implementation point of
view to track entropy rather than internal energy. However, at the level of
the equation of motion, this makes no difference, as this is
essentially just a choice of co-ordinate system.

This naturally leads the Pressure-Entropy variant (i.e. as opposed to
Pressure-Energy) of the Pressure-based schemes to be frequently chosen; here
the main smoothed quantity is pressure, with entropy being the
thermodynamic variable.

The Pressure-Entropy and Pressure-Energy scheme perform equally well on
hydrodynamics tests (see \citet{Hopkins2013} for a collection), but when
coupling to sub-grid physics there are some key differences.

For an entropy-based scheme, energy injection naturally leads to a conversion
between the requested energy input and an increase in entropy for the
relevant particle. Considering a Density-Entropy scheme to begin with
\citep[e.g.][]{Springel2002}, with only a smooth density $\hat{\rho}$,
\begin{align}
    P_i = (\gamma - 1) u_i \hat{\rho}_i,
\end{align}
with $P$ the pressure from the equation of state, $\gamma$ the ratio of specific
heats, and $u_i$ the particle energy per unit mass. In addition, the expression for
the pressure as a function of the entropy $A_i$,
\begin{align}
    P_i = A_i \hat{\rho}^\gamma.
\end{align}
Given that these should give the same thermodynamic pressure, the pressure
variable can be eliminated to give
\begin{align}
    u_i = \frac{A_i \hat{\rho}^{\gamma - 1}}{\gamma - 1}
\end{align}
and as these variables are independent for a change in energy $\Delta u$ the
change in entropy can be written
\begin{align}
    \Delta A_i = (\gamma - 1)\frac{\Delta u_i}{\hat{\rho}^{\gamma - 1}}.
\end{align}
For any energy based scheme (either Density-Energy or Pressure-Energy), it is
possible to directly modify the internal energy per unit mass $u$ of a
particle, and this directly corresponds to the same change in total energy of
the field. This is clearly also true here too for the Density-Entropy scheme.
Then, the sum of all energies (converted from entropies in the
Density-Entropy case) in the box will be the original value plus the injected
energy, without the requirement for an extra loop over
neighbours\footnote{This is only true given that the values entering the
smooth quantities, here the density, are not changed at the same time. In
practice, the mass of particles in cosmological simulations either does not
change or changes very slowly with time (due to sub-grid stellar enrichment
models for instance).}.

Now considering Pressure-Entropy, the smoothed pressure shown in Equation
\ref{eqn:smoothedpressure} at a particle depends on a
smoothed entropy over all of its neighbours. To connect the internal energy
and entropy of a particle again the equation of state can be used by
introducing a new variable, the weighted density $\bar{\rho}$,
\begin{align}
    \hat{P}_i = (\gamma - 1) u_i \bar{\rho}_i = A_i \bar{\rho}_i^\gamma
\end{align}
now being rearranged to eliminate the weighted density $\bar{\rho}$ such that
\begin{align}
    A_i(u_i) = \hat{P}_i^{1 - \gamma} ( \gamma - 1) u_i^\gamma,
    \label{eqn:Aasfuncu}
\end{align}
\begin{align}
    u_i(A_i) = \frac{A_i^{1/\gamma} \hat{P}^{1 - 1/\gamma}}{\gamma - 1}.
    \label{eqn:uasfuncA}
\end{align}

To inject energy into the \emph{field} by explicitly heating a single particle
$i$ in any entropy-based scheme the key is to find $\Delta A_i$ for a given
$\Delta u_i$. In a pressure-based scheme this is problematic, as (converting
Equation \ref{eqn:Aasfuncu} to a set of differences),
\begin{equation}
    \Delta A = \hat{P}_i(A_i)^{1 - \gamma} (\gamma - 1)
    (u_i + \Delta u)^\gamma - A_i,
    \label{eqn:deltaAasfuncu}
\end{equation}
to find this difference requires conversion via the smoothed pressure which
directly depends on the value of $A_i$. This also occurs for the particles
that neighbour $i$, meaning that there will be a non-zero change in the
energy $u_j$ that they report. Hence, this means that simply solving a linear
equation for $\Delta A(\Delta u)$ is not enough; whilst this can be
calculated, the true change in energy of the whole field will not be $\Delta
u$ (as it was in Density-Entropy) because of the changing pressures of the
neighbours. When attempting to inject energy it is vital that these
contributions to the total field energy are considered. To correctly
account for these changes, we must turn to an iterative solution.

\begin{figure}
    \centering
    \includegraphics{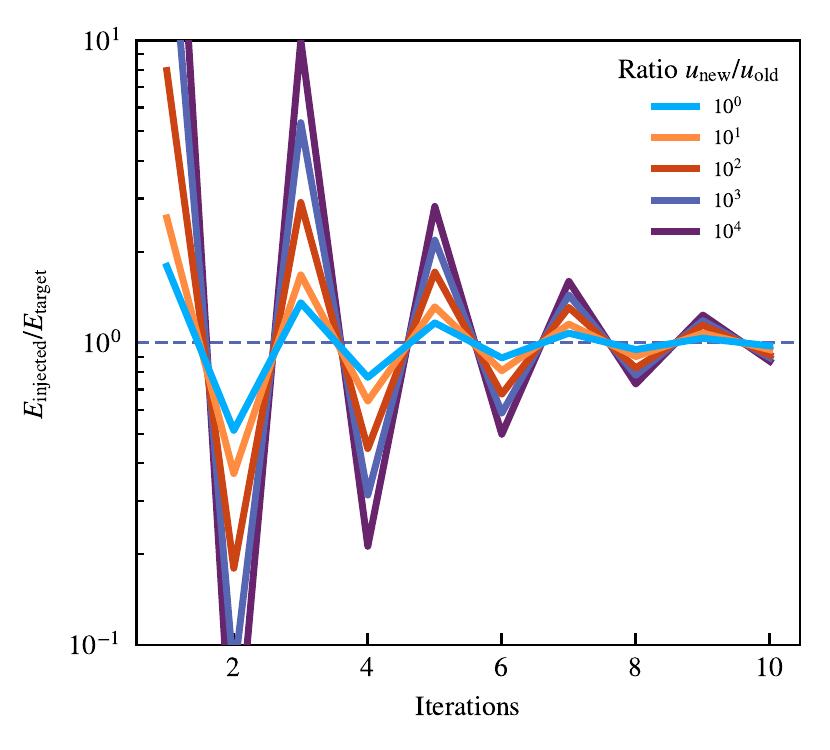}
    \caption{Energy injection as a function of iterations of the neighbour
    loop-based algorithm in Pressure-Entropy. Different coloured lines show
    ratios of injected energy to the original energy of the chosen particle,
    increasing in steps of 10. This algorithm allows for the correct energy
    to be injected into each particle after around 10 iterations, however
    more complex convergence criteria could be incorporated. A better
    estimate of the change in the smoothed pressure $\hat{P}$ could also
    significantly improve convergence.}
    \label{fig:energy_injection_better}
\end{figure}
\begin{figure}
    \centering
    \includegraphics{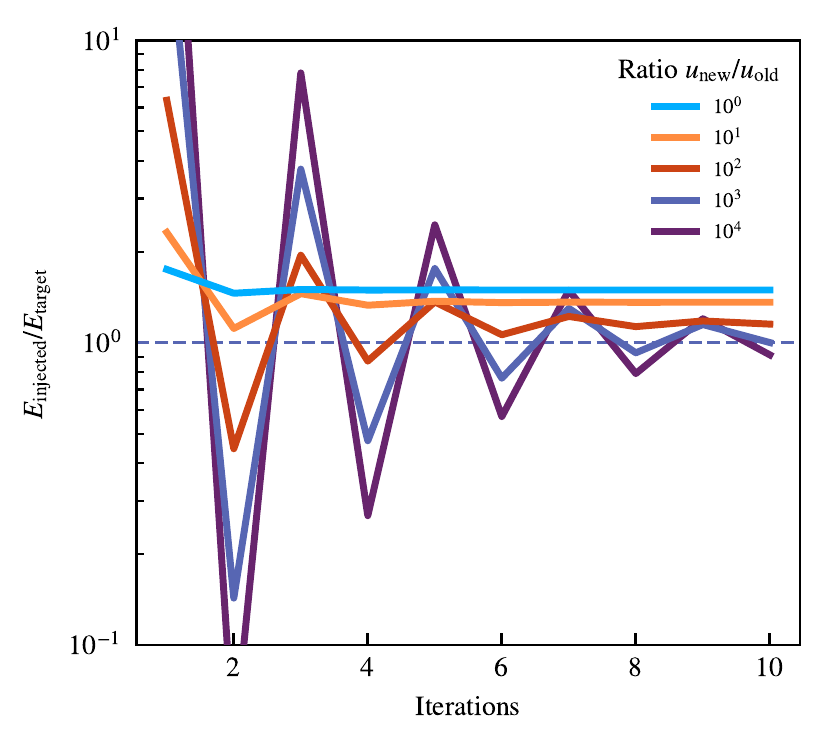}
    \caption{The same as Fig. \ref{fig:energy_injection_better}, however this
    time using an approximate algorithm that only updates the
    self-contribution of the heated particle. This version of the algorithm
    shows non-convergent behaviour at low energy injection values, but is
    significantly computationally cheaper than solutions that require
    neighbour loops during the iteration procedure.}
    \label{fig:energy_injection_EAGLE}
\end{figure}

A simple algorithm for injecting energy $\Delta u$ in this case would be as
follows:
\begin{enumerate}
    \item Calculate the total energy of all particles that neighbour
          the one that will have energy injected,
          $u_{{\rm field}, i} = \sum_j u(A_j, \hat{P}_j)$\footnote{More specifically
          we actually require all particles $j$ that see particle $i$ as a neighbour
          (rather than all particles $j$ that $i$ sees as a neighbour), which may be
          different in regions where the smoothing length varies significantly
          over a kernel, but this detail is omitted from the main discussion
          for clarity.}. \vspace{1.5mm}
    \item Find a target energy for the field, $u_{{\rm field}, t} = 
          u_{{\rm field}, i} + \Delta u$. \vspace{1.5mm}
    \item While the energy of the field $u_{{\rm field}} = \sum_j u(A_j, \hat{P}_j)$
          is outside of the bounds of the target energy:
    \begin{enumerate}
        \item Calculate $A_{\rm inject} = A(u_{{\rm field}, t} - u_{{\rm field}},
              \hat{P})$ for the particle that will have energy injected
              (i.e. apply Equation \ref{eqn:deltaAasfuncu} assuming that
              $\hat{P}_i$ does not change).\vspace{1.5mm}
        \item Add on $A_{\rm inject}$ to the entropy of the chosen particle.\vspace{1.5mm}
        \item Re-calculate the smoothed pressures for all neighbouring particles.\vspace{1.5mm}
        \item Re-calculate the energy of the field $u_{{\rm field}}$ (i.e. go to
              item \emph{iii} above).
    \end{enumerate}
\end{enumerate}
The results of this process, for various injection energies, are shown in
Fig. \ref{fig:energy_injection_better}. After around 10 iterations, the
requested injection of energy is reached. This process is valid only
for working on a single particle at a time, however, and as such would be
non-trivial to parallelise without the use of locks on particles that were
currently being modified. Suddenly changing the energy of a neighbouring
particle while this process was being performed would destroy the convergent
behaviour that is demonstrated in Fig. \ref{fig:energy_injection_better}.

Even without locks, this algorithm is computationally expensive, with many
thousands of operations required to change a single variable. Re-calculating
the smoothed pressure (step \emph{c}) for every particle multiple times per
step, is generally infeasible as it would require many thousands of operations
per particle per step. An ideal algorithm would not require neighbour
loops; only updating the self contribution for the heated
particle\footnote{This algorithm was implemented in the original \eagle{}
code using the weighted density, $\bar{\rho}$ as the smoothed quantity,
however this algorithm has been re-written to act on the smoothed pressure
for simplicity. See Appendix A1.1 of \citet{Schaye2015} for more details.}:
\begin{enumerate}
    \item Calculate the total energy of the particle that will have
          the energy injected, $u_{i, {\rm initial}} = u(A_i, \hat{P}_i)$.\vspace{1.5mm}
    \item Find a target energy for the particle, $u_{i, {\rm target}} = 
          u_{i, {\rm initial}} + \Delta u$. \vspace{1.5mm}
    \item While the energy of the particle $u_i = u(A_i, \hat{P}_i)$ is
          outside of the bounds of the target energy (tolerance here is $10^{-6}$,
          and is rarely reached) and the number of iterations is below the
          maximum (10):
    \begin{enumerate}
        \item Calculate $A_{\rm inject} = A(u_{i, t} - u_i,
              \hat{P})$ for the particle that will have the energy injected.\vspace{1.5mm}
        \item Add on $A_{\rm inject}$ to the entropy of that particle.\vspace{1.5mm}
        \item Update the self contribution to the smoothed pressure for 
              the injection particle by
              $\hat{P}_{i, {\rm new}} = \left[ \hat{P}_{i, {\rm old}}^{1/\gamma} +
              (A_{\rm new}^{1 / \gamma} - A_{\rm old}^{1 / \gamma})W_0 \right]^\gamma$
              with $W_0=W(0, h_i)$ the kernel self-contribution term.\vspace{1.5mm}
        \item Re-calculate the energy of the particle $u_i = u(A_i, \hat{P}_i)$
              using the new entropy and energy of that particle (i.e. go to \emph{iii}
              above).
    \end{enumerate}
\end{enumerate}
The implementation of the faster procedure is shown in Fig.
\ref{fig:energy_injection_EAGLE}. This simple algorithm leads to
significantly higher than expected energy injection for low (relative) energy
injection events. For the case of the requested energy injection being the
same as the initial particle energy, over 50\% too much energy is injected
into the field. For events that inject more entropy into particle $i$, the
value $A_i^{1/\gamma} W_{ij}$ for all neighbouring kernels becomes the
leading component of the smoothed pressure field. This allows the pressure
field to be dominated by this one particle, meaning that changes in
$A_i^{1/\gamma}$ represent linear changes in the pressures of neighbouring
particles, and hence allowing the simple methodology to correctly predict the
changes in the global internal energy field.

\begin{figure}
    \centering
    \includegraphics{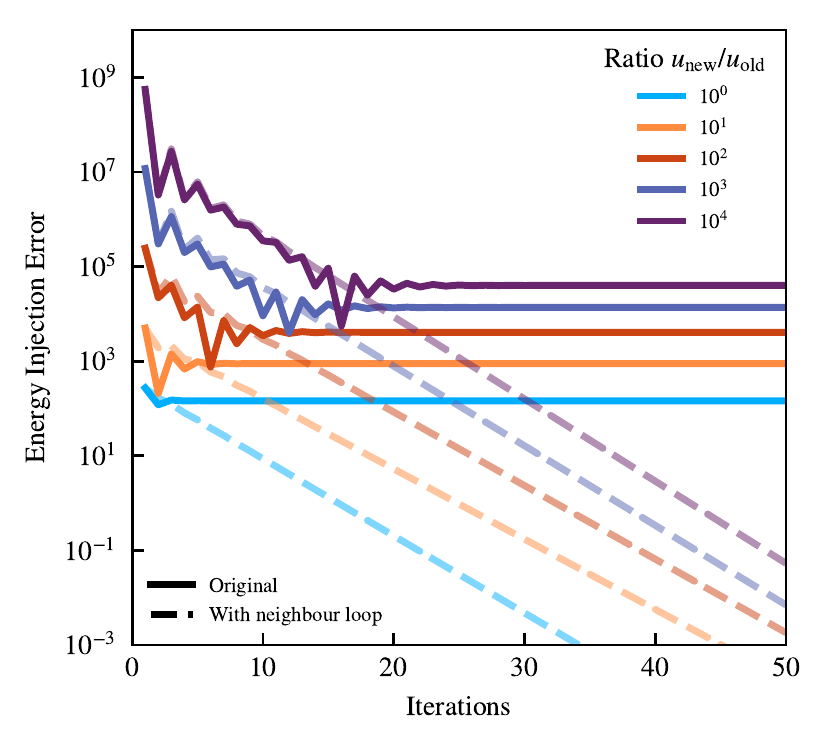}
    \caption{Comparison between the simple energy injection
    procedure (Fig. \ref{fig:energy_injection_EAGLE}, solid lines) against
    the method including a neighbour loop each iteration (Fig.
    \ref{fig:energy_injection_better}, dashed lines) for various energy
    injection values. The vertical axis here shows the energy offset from the
    true requested energy (in absolute arbitrary code units). The neighbour loop
    approach allows for the injected energy error to decrease with each
    iteration, where the simple procedure has a fixed (injection
    dependent) energy error that is reached rapidly at low values of energy
    injection where the entropies of neighbouring particles remain dominant.}
    \label{fig:energy_injection_compare}
\end{figure}

The error in the computationally cheaper injection method is
directly compared against the neighbour loop procedure from Fig. 
\ref{fig:energy_injection_better} in Fig. \ref{fig:energy_injection_compare}.
The extra energy injected per event is clear here; the method using
a full neighbour loop each iteration manages to reduce the error each iteration,
with the non neighbour loop method showing a fixed offset after a few iterations.
This also shows that the energy injection error grows as the amount injected
grows, despite this becoming a lower relative fraction of the requested
energy.

It is unclear exactly how much these errors impact the results of a full
cosmological run. For the case of supernovae following
\citet{DallaVecchia2012}, which has a factor of $u_{\rm new} / u_{\rm old}
\approx 10^4$ this should not represent a significant overinjection (the
energy converges within 10 iterations to around a percent or so). For
feedback pathways that inject a relatively smaller amount of energy (for
instance SNIa, AGN events on particles that have been recently heated, events
on particles in haloes with a high virial temperature, or schemes that inject
using smaller steps of energy or into multiple particles simultaneously)
there will be a significantly larger amount of energy injected than initially
expected. This uncontrolled energy injection is clearly undesirable.

\subsection{A Different Injection Procedure}

\begin{figure}
	\centering
	\includegraphics{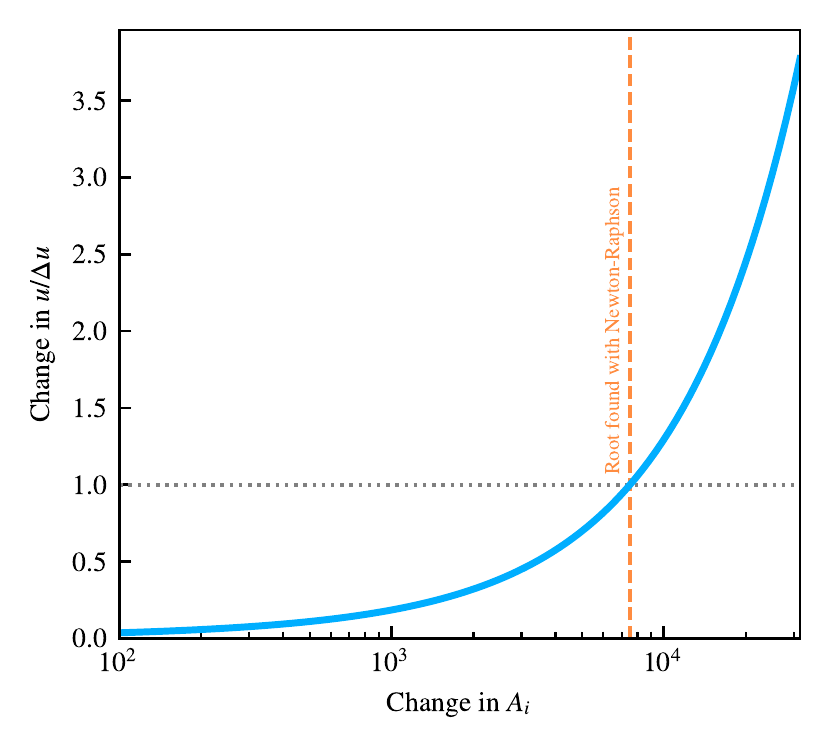}
	\caption{The blue line shows the dependence of change in
	field energy $u$ as a function of the change in the entropy $A_i$
	of a single particle, for a requested change in energy $\Delta
	u$. This change in energy $\Delta u$ corresponds to a heating
	event from $10^{3.5}$ K to $10^{7.5}$ K (a factor of $10^{4}$ in
	$u$), which corresponds to a typical energetic supernovae
	feedback event. The orange dashed line shows the predicted change
	in $A_i$ for this change $\Delta u$ from the iterative solution
	(using the Newton-Raphson method) of Equation
	\ref{eqn:energyinjectionsolveiteratively}.}
	\label{fig:betterthaneagle}
\end{figure}

Pressure-Entropy based schemes have been shown to be unable to inject the
correct amount of energy using a simple algorithm based on updating only a
single particle (i.e. without neighbour loops), however it is possible to
perform this task exactly within a single step by using an iterative solver
to find the change in entropy $\Delta A$.

To inject a set amount of energy $\Delta u$ the total energy of the field
$u_{\rm tot}$ must be modified by changing the properties of particle
$i$ (with neighbouring particles $j$), with
\begin{equation}
	u_{\rm tot} = \frac{1}{\gamma - 1}
                  \left( \sum_{j} 
                  m_j A_j^{1/\gamma}W_{ij}
                  \right)^{\gamma - 1}.
\end{equation}
This can be re-arranged to extract components specifically dependent
on the injection particle $i$,
\begin{align}
	u_{\rm tot} = \frac{1}{1 - \gamma} \sum_{j \neq i} & A_j^{1 / \gamma} \left(
		{p}_{j, i} + m_i A_i^{1 / \gamma} W_{ij}
	\right)^{\gamma - 1} \nonumber \\
	+ & A_i^{1 / \gamma} \left(
		{p}_{i, i} + m_i A_i^{1 / \gamma} W_{ii}
	\right)^{\gamma - 1},
\end{align}
with
\begin{equation}
	{p}_{a,b} = \hat{P}_b - m_b A^{1/\gamma}_b W_{ab}.
	\label{eqn:pab}
\end{equation}
Finally, now considering a change in energy $\Delta u$ as a function of
the change in entropy for particle $i$, $\Delta A$,	
\begin{align}
	\Delta u = \frac{1}{1 - \gamma} \sum_{j \neq i} & A_j^{1 / \gamma} \left(
		{p}_{j, i} + m_i (A_i + \Delta A)^{1 / \gamma} W_{ij}
	\right)^{\gamma - 1} \nonumber \\
	+ & (A_i + \Delta A)^{1 / \gamma} \left(
		{p}_{i, i} + m_i (A_i + \Delta A)^{1 / \gamma} W_{ii}
	\right)^{\gamma - 1} \nonumber \\ 
	- & u_{\rm tot},
	\label{eqn:energyinjectionsolveiteratively}
\end{align}
which can be solved iteratively using, for example, the Newton-Raphson
method. This method converges very well in just a few steps to calculate
the change in entropy $\Delta A$ as demonstrated in Fig. \ref{fig:betterthaneagle}.
In practice, this method would require two loops over the
neighbours of particle $i$ per injection event. In the first loop,
the values of $p_{j, i}$ and $W_{ij}$ would be calculated and stored,
with the iterative solver then used to find the appropriate value of
$\Delta A$. These changes would then need to be back-propagated to the
neighbouring particles, as their smoothed pressures $\hat{P}_j$ will
have changed significantly, reversing the procedure in Equation
\ref{eqn:pab}.

Such a scheme could potentially make a Pressure-Entropy based SPH method
viable for a model that uses energy injection. This procedure requires tens
of thousands of operations per thermal injection event, however, and as such
would require significant effort to implement efficiently.

This also highlights a possible issue with Pressure-Energy based SPH schemes,
as even in this case, where it is much simpler to make changes to the global
energy field, changes to the internal energy of a particle must be
back-propagated to neighbours to ensure that the pressure and internal energy
fields remain consistent. These errors also compound, should more than one
particle in a kernel be heated without the back-propagation of changes.

\section{Equations of Motion}
\label{sec:eom}

So far only static fields have been under consideration; before moving on to
discussing the effects of sub-grid cooling on pressure-based schemes, the
\emph{dynamics} part of SPH must be considered. Below only two equations of
motion are described, the one corresponding to Density-Energy, and the
equation of motion for Pressure-Energy SPH. For a more expanded derivation of
the following from a Lagrangian and the first law of Thermodynamics see
\citet{Hopkins2013}, or the \swift{} simulation code theory
documentation\footnote{ \url{http://www.swiftsim.com}}.

\subsection{Density-Energy}

For Density-Energy the smoothed quantity of interest is the smoothed
mass density (Equation \ref{eqn:sphdensity}). This leads to a corresponding
equation of motion for velocity of
\begin{align}
    \frac{\mathrm{d}\mathbf{v}_i}{\mathrm{d}t} = 
    - \sum_j m_j &\left[
        f_i \frac{P_i}{\hat{\rho}_i^2} \nabla W(r_{i\!j}, h_i) +
        f_j \frac{P_j}{\hat{\rho}_j^2} \nabla W(r_{ji}, h_j)
    \right],
    \label{eqn:density_energy_v_eom}
\end{align}
with the $f_i$ here representing correction factors for interactions
between particles with different smoothing lengths
\begin{equation}
    f_i = \left(1 + \frac{h_i}{n_d \hat{\rho}_i} \frac{\partial
    \hat{\rho}_i}{\partial h_i}\right)^{-1}.
    \label{eqn:density_energy_f_fac}
\end{equation}
This factor also enters into the equation of motion for the internal
energy
\begin{equation}
    \frac{\mathrm{d}u_i}{\mathrm{d}t} = \sum_j m_j f_i \frac{P_i}{\hat{\rho}_i^2}
    \mathbf{v}_{ij} \cdot \nabla W(r_{ij}, h_i).
    \label{eqn:density_energy_u_eom}
\end{equation}

\subsection{Pressure-Energy}

For Pressure-Energy SPH, the thermodynamic quantity $u$ remains the same as for
Density-Energy, but the smoothed pressure field $\hat{P}$ is introduced (see
Equation \ref{eqn:smoothedpressure}). This is then used in the equation of
motion for the particle velocities
\begin{equation}
    \frac{\mathrm{d} \mathbf{v}_i}{\mathrm{d} t} = -\sum_j (\gamma - 1)^2 m_j
    u_j u_i 
    \left[ \frac{f_{ij}}{\hat{P}_i} \nabla W(r_{ij}, h_i) +
    \frac{f_{ji}}{\hat{P}_j} \nabla W(r_{ji}, h_j)   \right].
    \label{eqn:pressure_energy_v_eom}
\end{equation}
with the $f_{ij}$ now depending on both particle $i$ and $j$
\begin{equation}
    f_{ij} = 1 - \left[\frac{h_i}{n_d (\gamma - 1) \hat{n}_i m_j u_j}
             \frac{\partial \hat{P}_i}{\partial h_i} \right]
             \left( 1 + \frac{h_i}{n_d \hat{n}_i}
             \frac{\partial \hat{n}_i}{\partial h_i} \right)^{-1},
    \label{eqn:pressure_energy_f_fac}
\end{equation}
with $\hat{n}$ the local particle number density (Equation
\ref{eqn:numberdensity}). Again, this factor enters into the equation of
motion for the internal energy
\begin{equation}
    \frac{\mathrm{d} u_i}{\mathrm{d} t} = (\gamma - 1)^2 \sum_j m_j
    u_i u_j \frac{f_{ij}}{\hat{P}_i} \mathbf{v}_{ij} \cdot \nabla W_{ij}.
    \label{eqn:pressure_energy_u_eom}
\end{equation}

\subsection{Choosing an Appropriate Time-Step}

To integrate these forward in time, an appropriate time-step between the
evaluation of these smoothed equations of motion must be chosen.  SPH schemes
typically use a modified version of the Courant–Friedrichs–Lewy \citep[CFL,
][]{Courant1928} condition to determine this step length. The CFL condition
takes the form of
\begin{equation}
    \Delta t = C_{\rm CFL} \frac{H_i}{c_{\rm s}},
    \label{eqn:cfl}
\end{equation}
with $c_{\rm s}$ the local sound-speed, and $C_{\rm CFL}$ a constant that
should be strictly less than 1.0, typically taking a value of 0.1-0.3\footnote{In
practice this $c_s$ is usually replaced with a signal velocity $v_{\rm sig}$
that depends on the artificial viscosity parameters. As the implementation of
an artificial viscosity is not discussed here, this detail is omitted for
simplicity.}. Computing this sound-speed is a simple affair in density-based
SPH, with it being a particle-carried property that is a function solely of
other particle carried properties,
\begin{equation}
    c_{\rm s} = \sqrt{\gamma \frac{P}{\hat{\rho}}} = \sqrt{\gamma (\gamma - 1) u}.
    \label{eqn:speed_of_sound_density}
\end{equation}
For pressure-based schemes this requires a little more thought. The same
sound-speed can be used, but this is not representative of the variables that
actually enter the equation of motion. To clarify this, first consider the
equation of motion for Density-Energy (Equation \ref{eqn:density_energy_v_eom})
and re-write it in terms of the sound-speed,
\begin{align}
    \frac{\mathrm{d}\mathbf{v}_i}{\mathrm{d} t} \sim \frac{c_{{\rm s}, i}^2}{\hat{\rho}_i}
\nabla_i W_{ij},
  \nonumber
\end{align}
and for Pressure-Energy (Equation \ref{eqn:pressure_energy_v_eom})
\begin{align}
  \frac{\mathrm{d}\mathbf{v}_i}{\mathrm{d} t} \sim (\gamma - 1)^2
  \frac{u_i u_j}{\hat{P}_i} \nabla_i W_{ij}.
  \nonumber
\end{align}
From this it is reasonable to assume that the sound-speed, i.e. the speed at
which information propagates in the system through pressure waves, is given by
the expression
\begin{align}
    c_{\rm s} = (\gamma - 1) u_i \sqrt{\gamma \frac{\hat{\rho}_i}{\hat{P}_i}}.
    \label{eqn:pressure_energy_wrong_soundspeed}
\end{align}
This expression is dimensionally consistent with a sound-speed, and includes
the gas density information (through $\hat{\rho}$), traditionally used for
sound-speeds, as well as including the extra information from the smoothed
pressure $\hat{P}$. However, such a sound-speed leads to a considerably
\emph{higher} time-step in front of a shock wave (where the smoothed pressure
is higher, but the smooth density is relatively constant), leading to time integration
problems. Using
\begin{align}
   c_{\rm s} = \sqrt{\gamma \frac{\hat{P}_i}{\hat{\rho}_i}}
  \label{eqn:pressure_energy_soundspeed}
\end{align}
instead of Equation \ref{eqn:pressure_energy_wrong_soundspeed} leads to a
sound-speed that does not represent the equation of motion as directly
but does not lead to time-integration problems, and effectively represents
a smoothed internal energy field. It is also possible to use the same
sound-speed using the particle-carried internal energy directly above.

\section{Time Integration}
\label{sec:timeintegration}

A typical astrophysics SPH code will use Leapfrog integration or a
velocity-verlet scheme to integrate particles through time \citep[see e.g.
][]{Hernquist1989,Springel2005,Borrow2018}. This approach takes the
accelerations, $\mathbf{a}_i = \mathrm{d}\mathbf{v}_i / \mathrm{d} t$, and the
velocities, $\mathbf{v}_i = \mathrm{d}\mathbf{r}_i / \mathrm{d} t$ and solves
the system for the positions $r_i(t)$ as a function of time.  It is convenient
to write the equations as follows (for each particle):
\begin{align}
    \mathbf{v}\left(t + \frac{\Delta t}{2}\right) & =
        \mathbf{v}(t) + \frac{\Delta t}{2}\mathbf{a}(t),\\
    \mathbf{r}\left(t + \Delta t\right) & = 
        \mathbf{r}(t) + \mathbf{v}\left(t + \frac{\Delta t}{2}\right)\Delta t, \\
    \mathbf{v}\left(t + \Delta t\right) & = \mathbf{v}
        \left(t + \frac{\Delta t}{2}\right) + \frac{\Delta t}{2}\mathbf{a}(t + \Delta t),
    \label{eqn:KDK}
\end{align}
commonly referred to (in order) as a Kick-Drift-Kick scheme. Importantly,
these equations must be solved for all variables of interest.

This leapfrog time-integration is prized for its second order accuracy (in
$\Delta t$) despite only including first order operators, due to cancelling
second order terms as well as its manifest conservation of energy \citep{
Hernquist1989}.

\subsection{Multiple Time-Stepping}

As noted above, it is possible to find a reasonable time-step to
evolve a given hydrodynamical system with using the CFL condition
(Equation \ref{eqn:cfl}). This condition applies on a
particle-by-particle basis, meaning that to evolve the whole
\emph{system} a method for combining these individual time-steps into
a global mechanism must be devised. In less adaptive problems than
those considered here (e.g. those with little dynamic range in
smoothing length), it is reasonable to find the minimal time-step over
all particles, and evolve the whole system with this time-step. This
scenario is frequently referred to as `single-d$t$'.

For a cosmological simulation, however, the huge dynamic range in
smoothing length (and hence time-step) amongst particles means that
evolving the whole system with a single time-step would render most
simulations infeasible \citep{Borrow2018}. Instead, each particle
is evolved according to its own time-step (referred to as a
multi-d$t$ simulation) using a so-called `time-step hierarchy'
as originally described in \citet{Hernquist1989}. This choice
is common-place in astrophysics codes \citep{Teyssier2002, 
Springel2005}.

In some steps in a multi-d$t$ simulation only the particles on
the very shortest time-steps are updated in a loop over their
neighbours to re-calculate, for example, $\hat{\rho}$ (referred to as these
particles being `active'). The rest of the particles
are referred to as being `inactive'. As the inactive particles may
interact with the active ones, their properties must be interpolated,
or drifted, to the current time. 

For particle-carried quantities, such as the internal energy $u$,
a simple first-order equation is used,
\begin{equation}
    u\left(t + \Delta t\right) = 
        u + \frac{\mathrm{d} u}{\mathrm{d} t}\Delta t.
    \label{eqn:internal_energy_drift}
\end{equation}

\subsection{Drifting Smoothed Quantities}
\label{sec:driftoperators}

As a particle may experience many more drift steps than loops over neighbours
(that are only performed for active particles), it is important to have drift
operators ($\mathrm{d}\hat{x} / \mathrm{d}t$) for smoothed quantities
$\hat{x}$ to interpolate their values between full time-steps. This is
achieved through taking the time differential of smoothed quantities.
Starting with the simplest, the smoothed number density,
\begin{align}
    \frac{\mathrm{d}\hat{n}_i}{\mathrm{d}t} &=
    \sum_j \frac{\mathrm{d} W(r_{ij}, h_i)}{\mathrm{d}t}, \nonumber \\
    &= \sum_j \mathbf{v}_{ij} \cdot \nabla_j W(r_{ij}, h_i).
\end{align}
Following this process through for the smoothed quantities of interest yields
\begin{align}
    \frac{\mathrm{d}\hat{\rho}_i}{\mathrm{d}t} &= 
        \sum_j m_j \mathbf{v}_{ij} \cdot \nabla_j W(r_{ij}, h_i),\\
    \frac{\mathrm{d}\hat{P}_i}{\mathrm{d}t} &= 
        (\gamma - 1)\sum_j m_j\left(W_{ij}\frac{\mathrm{d}u_j}{\mathrm{d}t} +
        u_j \mathbf{v}_{ij} \cdot \nabla_j W_{ij}\right),
    \label{eqn:drift_P}
\end{align}
for the smoothed density and pressure respectively, with $W_{ij} = W(r_{ij},
h_i)$. In the smoothed density case, the pressure is re-calculated at each
drift step from the now drifted internal energy and density using the
equation of state\footnote{Note that the first equation for the smoothed
density corresponds to the SPH discretisation of the continuity equation
\citep{Monaghan1992}, but the second equation makes little physical
sense.}.

The latter drift equation, due to its inclusion of $\mathrm{d}u_j/\mathrm{d}t$
(i.e. the rate of change of internal energy of all neighbours of particle
$i$), presents several issues. This sum
is difficult to compute in practice; it requires that all of the
$\mathrm{d}u_j/\mathrm{d}t$ are set before a neighbour loop takes place.
This would require an extra loop over neighbours after the `force' loop,
which has generally been considered computationally infeasible for a scheme
that purports to be so cheap. In practice, the following is used to drift 
the smoothed pressure:
\begin{equation}
	 \frac{\mathrm{d}\hat{P}_i}{\mathrm{d}t} =
	 	\frac{\mathrm{d}\hat{\rho}_i}{\mathrm{d}t} \cdot
	 	\frac{\mathrm{d}u_i}{\mathrm{d}t}~,
	 \label{eqn:drift_P_with_rho}
\end{equation}
which clearly does not fully capture the expected behaviour of Equation
\ref{eqn:drift_P} as it only includes the rate of change of the internal
energy for particle $i$, discarding the contribution from neighbours.

Such behaviour becomes particularly problematic in cases where sub-grid
cooling is used, where particles within a kernel may have both very large
$\mathrm{d}u_j/\mathrm{d}t$ (where $(\mathrm{d}u_j/\mathrm{d}t)\Delta t$ is
comparable to $u_j$), and $\mathrm{d}u_j/\mathrm{d}t$ that vary rapidly with
time. Consider the case where an active particle cools rapidly from some
temperature to the equilibrium temperature in one step (which occurs
frequently in a typical cosmological simulation where no criterion on the
time-step for $\mathrm{d}u/\mathrm{d}t$ is included to ensure the number of
steps required to complete the calculation remains reasonable whilst
employing implicit cooling). If this particle has a neighbour at the
equilibrium temperature that is inactive, the pressure for the neighbouring
particle will remain significantly (potentially orders of magnitude) higher
than what is mandated by the local internal energy field, leading to force
errors of a similar level.

To apply these drift operators to smoothed quantities, instead of using a
linear drift as in Equation \ref{eqn:internal_energy_drift}, the analytic
solution to these first order differential equations is used. For a smooth
quantity $\hat{x}$ it is drifted forwards in time using
\begin{equation}
   \hat{x}(t + \Delta t) = \hat{x}(t)\cdot \exp\left(
      \frac{1}{\hat{x}} \frac{\mathrm{d}\hat{x}}{\mathrm{d}t}
   \right).
   \label{eqn:smooth_drift}
\end{equation}
This also has the added benefit of preventing the smoothed quantities from
becoming negative. For this to be accurate, it requires an accurate
$\mathrm{d}\hat{x}/\mathrm{d}t$ term.

\subsection{Impact of Drift Operators in multi-d$t$}

\begin{figure}
    \centering
    \includegraphics{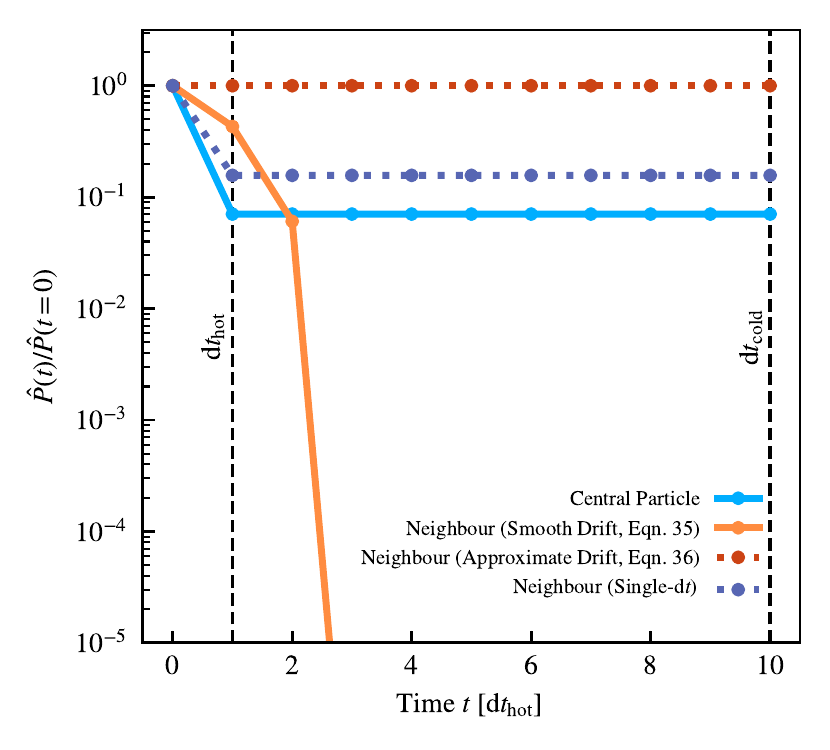}
    \caption{Smooth pressure as a function of time for different
    strategies in a uniform fluid of `cold' particles, with one
    initially `hot' particle with a temperature 100 times higher
    than the cold particles that cools to the `cold' temperature
    in one time-step. The solid blue line shows the pressure
    of the central particle as a function of time (relative to
    its initial pressure). The dashed blue line shows the pressure
    of the closest `hot' neighbour in a single-d$t$ scenario, i.e.
    the whole system is evolved with time-step d$t_{\rm hot}$. This
    shows the true answer for the pressure of the neighbour particle.
    The dotted red line shows the result of drifting the cold particle
    with Equation \ref{eqn:drift_P_with_rho}. As this particle has
    no cooling rate, and the fluid is stationary, the pressure does
    not change. The solid orange line shows the result of drifting
    using Equation \ref{eqn:drift_P}. This rapidly leads to the
    particle having a pressure of zero, a highly undesirable result.
    Note that the orange line does not follow the dashed blue line
    in the first few steps due to different drifting schemes for
    smoothed and particle-carried quantities (Equation
     \ref{eqn:internal_energy_drift} and \ref{eqn:smooth_drift}).}
    \label{fig:pressure_ratio_drift}
 \end{figure}
 
Whilst the true drift operator for $\hat{P}$ appears to be impractical from a
computational perspective due to the requirement of another loop over
neighbours, at first glance it appears that the use of this correct drift
operator would remedy the issues with cooling. Unfortunately, in a multi-d$t$
simulation where active and in-active particles are mixed, this `correct'
operator can still lead to negative pressures when applied.

In Fig. \ref{fig:pressure_ratio_drift} the different ways of drifting smooth
pressure in a multi-d$t$ simulation are explored. In this highly idealised
test, a cubic volume of uniform `cold' fluid is considered.  A single particle
at the center is set to have a `hot' temperature of 100 times higher than the
background fluid, and is set to have a cooling rate that ensures that it cools
to the `cold' temperature within its first time-step. This scenario is similar
to a hot $10^6$ K particle in the CGM cooling to join particles in the ISM at
the $10^4$ K equilibrium temperature. The difference between the time-step of
the hot and cold particles, implied by Equation \ref{eqn:cfl}, is a factor of
10 (when using the original definition of sound-speed, see Equation
\ref{eqn:speed_of_sound_density}).  Here the cold particle is drifted ten times
to interact with its hot neighbour over a single time-step of its own. In
practice, this scenario would evolve slightly differently, with the previously
hot particle having its time-step re-set to d$t_{\rm cool}$ after it has cooled
to the equilibrium temperature, but the nuances of the time-step hierarchy are
ignored here for simplicity.

\begin{figure}
\centering
\includegraphics{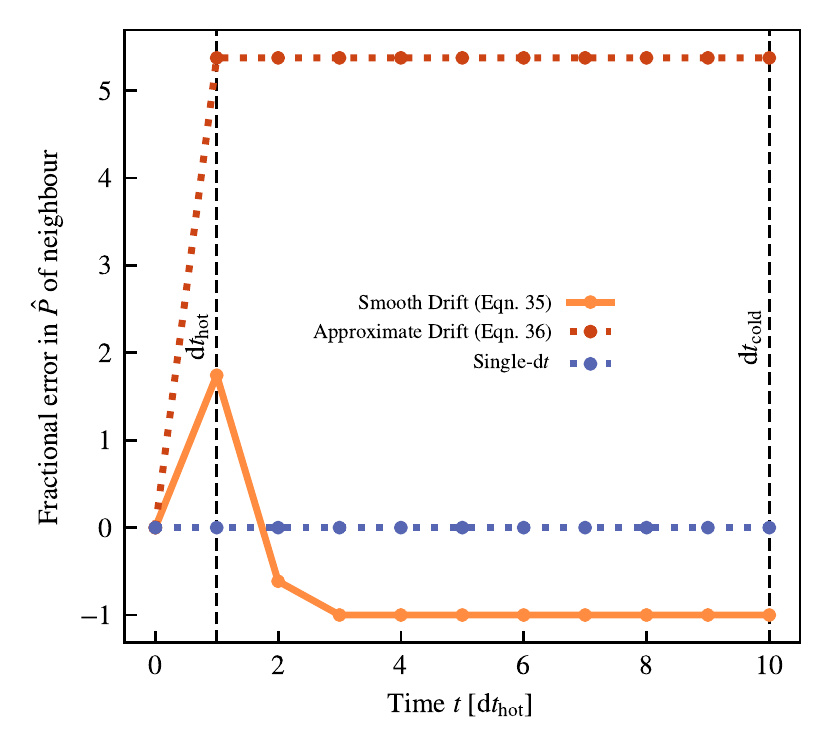}
\caption{The same lines as Fig. \ref{fig:pressure_ratio_drift},
except now showing the `error' as a function of time
relative to the single-d$t$ case (blue dashed line) of
the pressure $\hat{P}$ of the nearest neighbour to the
`hot' particle. Here the fractional error is defined as
$\hat{P}(t) - \hat{P}_{\rm{single-d}t} / \hat{P}_{\rm{single-d}t}$.
The orange line showing the drifting using Equation \ref{eqn:drift_P}
shows that the pressure rapidly drops to zero after around four
steps. The red dotted line (Equation \ref{eqn:drift_P_with_rho})
shows the offset in pressure that is maintained even after the
central `hot' particle cools.}
\label{fig:pressure_error_drift}
\end{figure}

The three drifting scenarios proceed very differently. In Fig.
\ref{fig:pressure_error_drift} the fractional errors relative to
the single-d$t$ case are shown.

In the case of the drift using
Equation \ref{eqn:drift_P}, the pressure rapidly
drops to zero. This is prevented from becoming negative thanks to
the integration strategy that is employed (Equation
\ref{eqn:smooth_drift}); the rate of $\mathrm{d}\hat{P}/\mathrm{d}t$
is high enough to lead to negative pressures within a few
drift steps should a simple linear integration strategy like
that employed for the internal energy (Equation \ref{eqn:internal_energy_drift})
be used. Because there is only a linear time integration (with a poorly
chosen time-step for the equation to be evolved) method for a now
non-linear problem (as there is a significant $\mathrm{d}^2 u / \mathrm{d}t^2$
from changes in cooling rate) errors naturally manifest.

The drift operator using a combination of the local 
cooling rate and density time differential (Equation \ref{eqn:drift_P_with_rho})
is the safest, leading to pressures that are higher than expected;
this does however come at the cost of larger relative errors in the
pressure (500\% increase v.s. 100\% decrease; both of these are 
highly undesirable).

\subsubsection{Limiting time-steps}

One way to address the issues presented in Fig. \ref{fig:pressure_ratio_drift}
is to limit the time-steps between neighbouring particles. Such a 
`time-step limiter' is common-place in galaxy formation simulations,
as they are key to capturing the energy injected during feedback
events \citep[see e.g.][]{Durier2012}. In addition, the use of
the `smoothed' sound-speed (from Equation \ref{eqn:pressure_energy_soundspeed})
ensures that the neighbouring particle has a time-step that is much
closer to the time-step of the `hot' particle than the sound-speed
based solely on the internal energy of each particle alone.
However, as Fig. \ref{fig:pressure_error_drift} shows, even only after
one intervening time-step (i.e. after $\mathrm{d}t_{\rm hot}$), there
is a 50\% to 500\% error in the pressure of the neighbouring particle.

This error in the pressure of the neighbouring particle
represents a poorly tracked non-conservation of energy. An incorrect
relationship between the local internal energy and pressure field
of the particles leads directly to force errors of the same magnitude.
Because of the conservative and symmetric structure of the applied
equations of motion, however, this does not lead to the total energy
of the fluid changing over time (i.e. the sum of the kinetic and
internal energy of the fluid remains constant), instead manifesting as
unstable dynamics. 

\section{Conclusions}
\label{sec:conclusions}

The Pressure-Energy and Pressure-Entropy schemes have been prized for their
ability to capture contact discontinuities significantly better than their
Density-based cousins due to their use of a directly smoothed pressure field
\citep{Hopkins2013}. However, there are several disadvantages to using these
schemes that have been presented:
\begin{itemize}
   \item Injecting energy in a Pressure-Entropy based scheme requires the use
         of an iterative solver and many transformations between variables. This
         makes this scheme computationally expensive, and as such for this to be
         used in practice an efficient implementation is required. Approximate
         solutions do exist, but result in incorrect amounts of energy being
         injected into the field when particles are heated only by a small
         amount (typically by less than 100 times their own internal energy).
         This occurs even in the case where the fluid is evolved with a single,
         global, time-step, and is complicated even further by the inclusion of
         the multiple time-stepping scheme that is commonplace in cosmological
         simulations.
   \item In a Pressure-Energy based scheme, the injection of energy in a
         multi-d$t$ simulation requires either `waking up' all of the neighbours
         of the affected particle (and forcing them to be active in the next
         time-step), or a loop over these neighbours to back-port changes to
         their pressure due to the changes in internal energy of the heated
         particle.  This is a computationally expensive procedure, and is
         generally avoided in the practical use of these schemes. As such, while
         no explicit energy conservation errors manifest, there is an offset
         between the energy field represented by the particle distribution and
         the associated smooth pressure field in practical implementations.
   \item These issues also manifest themselves in cases where energy is
         removed from active particles, such as an `operator-splitting'
         radiative cooling scheme where energy is directly removed from
         particles.
   \item Correctly `drifting' the smoothed pressure of particles (as is
         required in a multi-d$t$ simulation) requires knowing the time
         differential of the smoothed pressure. To compute this, either an extra
         loop over neighbours is required for active particles, or an
         approximate solution based on the time differential of the density
         field and internal energy field is used. This approximate solution does
         not account for the changes taking place in the local internal energy
         field and as such does not correctly capture the evolution of the
         smoothed pressure.
   \item Even when using the `correct' drift operator for the smoothed pressure
         significant pressure, and hence force, errors can occur when particles
         cool rapidly. This can be mitigated somewhat with time-step limiting
         techniques (either through the use of a time-step limiter like the one
         described in \citet{Durier2012} or through a careful construction of a
         more representative sound-speed) but it is not possible to prevent
         errors on the same order as the relative energy difference between
         the cooling particle and its neighbours.
\end{itemize}
All of the above listed issues are symptomatic of one main flaw in
these schemes; the SPH method assumes that the variables being
smoothed over vary slowly during a single time-step. This is often
true for the internal energy or particle entropy in idealised
hydrodynamics tests, but in practical simulations with sub-grid
radiative cooling (and energy injection) this leads to significant
errors. These errors could be mitigated by using a different cooling
model, where over a single time-step only small changes in the
energies of particles could be made (i.e. by limiting the time-steps
of particles to significantly less than their cooling time), however
this would render most cosmological simulations impractical to
complete due to the huge increase in the number of time-steps to
finish the simulation that this would imply.

Thankfully, due to the explicit connection between internal energy and
pressure in the Density-based SPH schemes, they do not suffer the same ills.
They also smooth over the mass field, which either does not vary or generally
varies very slowly (on much larger timescales than the local dynamical time).
As such, the only recommendation that it is possible to make is to move away
from Pressure-based schemes in favour of their Density-based cousins, solving
the surface tension issues at contact discontinuities with artificial
conduction instead of relying on the smoothed pressure field from
Pressure-based schemes. It is worth noting that most modern implementations
of the Pressure-based schemes already use an artificial conduction (also known as
energy diffusion) term to resolve residual errors in fluid mixing problems
\citet{Hu2014,Hopkins2015}. Of particular note is the lack of phase mixing
(due to the non-diffusive nature of SPH) between hot and cold fluids, even in
Pressure-SPH.

\section{Acknowledgements}
\label{sec:acknowledgements}

The authors would like to thank Joop Schaye and Claudio Dalla Vecchia for
their helpful conversations.
JB is supported by STFC studentship ST/R504725/1. 
MS is supported by the Netherlands Organisation for Scientific Research
(NWO) through VENI grant 639.041.749.
This work used the DiRAC@Durham facility managed by the Institute for
Computational Cosmology on behalf of the STFC DiRAC HPC Facility
(www.dirac.ac.uk). The equipment was funded by BEIS capital funding
via STFC capital grants ST/K00042X/1, ST/P002293/1, ST/R002371/1 and
ST/S002502/1, Durham University and STFC operations grant
ST/R000832/1. DiRAC is part of the National e-Infrastructure.

\subsection{Software Citations}

This paper made use of the following software packages:
\begin{itemize}
    \item {\tt python} \citep{VanRossum1995}, with the following libraries
    \begin{itemize}
    	\item {\tt numpy} \citep{Harris2020}
    	\item {\tt scipy} \citep{SciPy1.0Contributors2020}
        \item {\tt matplotlib} \citep{Hunter2007}
        \item {\tt numba} \citep{Lam2015}
        \item {\tt swiftsimio} \citep{Borrow2020a}
    \end{itemize}
\end{itemize}

\section{Data Availability}
\label{sec:dataavail}

No new data were generated or analysed in support of this research.
The figures in this paper were all produced from the equations
described within, and all programs used to generate those figures
are available in the {\tt GitHub} repository at
\url{https://github.com/JBorrow/pressure-sph-paper-plots}. They
are also available in the online supplementary material accompanying
this article.

\bibliographystyle{mnras}
\bibliography{bibliography}

\begin{thebibliography}{}
\makeatletter
\relax
\def\mn@urlcharsother{\let\do\@makeother \do\$\do\&\do\#\do\^\do\_\do\%\do\~}
\def\mn@doi{\begingroup\mn@urlcharsother \@ifnextchar [ {\mn@doi@}
  {\mn@doi@[]}}
\def\mn@doi@[#1]#2{\def\@tempa{#1}\ifx\@tempa\@empty \href
  {http://dx.doi.org/#2} {doi:#2}\else \href {http://dx.doi.org/#2} {#1}\fi
  \endgroup}
\def\mn@eprint#1#2{\mn@eprint@#1:#2::\@nil}
\def\mn@eprint@arXiv#1{\href {http://arxiv.org/abs/#1} {{\tt arXiv:#1}}}
\def\mn@eprint@dblp#1{\href {http://dblp.uni-trier.de/rec/bibtex/#1.xml}
  {dblp:#1}}
\def\mn@eprint@#1:#2:#3:#4\@nil{\def\@tempa {#1}\def\@tempb {#2}\def\@tempc
  {#3}\ifx \@tempc \@empty \let \@tempc \@tempb \let \@tempb \@tempa \fi \ifx
  \@tempb \@empty \def\@tempb {arXiv}\fi \@ifundefined
  {mn@eprint@\@tempb}{\@tempb:\@tempc}{\expandafter \expandafter \csname
  mn@eprint@\@tempb\endcsname \expandafter{\@tempc}}}

\bibitem[\protect\citeauthoryear{Agertz et~al.,}{Agertz
  et~al.}{2007}]{Agertz2007}
Agertz O.,  et~al., 2007, \mn@doi [Monthly Notices of the Royal Astronomical
  Society] {10.1111/j.1365-2966.2007.12183.x}, 380, 963

\bibitem[\protect\citeauthoryear{Borrow \& Borrisov}{Borrow \&
  Borrisov}{2020}]{Borrow2020a}
Borrow J.,  Borrisov A.,  2020, \mn@doi [Journal of Open Source Software]
  {10.21105/joss.02430}, 5, 2430

\bibitem[\protect\citeauthoryear{Borrow, Bower, Draper, Gonnet  \&
  Schaller}{Borrow et~al.}{2018}]{Borrow2018}
Borrow J.,  Bower R.~G.,  Draper P.~W.,  Gonnet P.,   Schaller M.,  2018,
  Proceedings of the 13th SPHERIC International Workshop, Galway, Ireland, June
  26-28 2018, pp 44--51

\bibitem[\protect\citeauthoryear{Courant, Friedrichs  \& Lewy}{Courant
  et~al.}{1928}]{Courant1928}
Courant R.,  Friedrichs K.,   Lewy H.,  1928, \mn@doi [Mathematische Annalen]
  {10.1007/BF01448839}, 100, 32

\bibitem[\protect\citeauthoryear{Cui et~al.,}{Cui et~al.}{2019}]{Cui2019}
Cui W.,  et~al., 2019, \mn@doi [Monthly Notices of the Royal Astronomical
  Society] {10.1093/mnras/stz565}, 485, 2367

\bibitem[\protect\citeauthoryear{Cullen \& Dehnen}{Cullen \&
  Dehnen}{2010}]{Cullen2010}
Cullen L.,  Dehnen W.,  2010, \mn@doi [Monthly Notices of the Royal
  Astronomical Society] {10.1111/j.1365-2966.2010.17158.x}, 408, 669

\bibitem[\protect\citeauthoryear{Dalla~Vecchia \& Schaye}{Dalla~Vecchia \&
  Schaye}{2012}]{DallaVecchia2012}
Dalla~Vecchia C.,  Schaye J.,  2012, \mn@doi [Monthly Notices of the Royal
  Astronomical Society] {10.1111/j.1365-2966.2012.21704.x}, 426, 140

\bibitem[\protect\citeauthoryear{Dav{\'e}, {Angl{\'e}s-Alc{\'a}zar}, Narayanan,
  Li, Rafieferantsoa  \& Appleby}{Dav{\'e} et~al.}{2019}]{Dave2019}
Dav{\'e} R.,  {Angl{\'e}s-Alc{\'a}zar} D.,  Narayanan D.,  Li Q.,
  Rafieferantsoa M.~H.,   Appleby S.,  2019, \mn@doi [Monthly Notices of the
  Royal Astronomical Society] {10.1093/mnras/stz937}, 486, 2827

\bibitem[\protect\citeauthoryear{Dehnen \& Aly}{Dehnen \&
  Aly}{2012}]{Dehnen2012}
Dehnen W.,  Aly H.,  2012, \mn@doi [Monthly Notices of the Royal Astronomical
  Society] {10.1111/j.1365-2966.2012.21439.x}, 425, 1068

\bibitem[\protect\citeauthoryear{Dolag, Borgani, Murante  \& Springel}{Dolag
  et~al.}{2009}]{Dolag2009}
Dolag K.,  Borgani S.,  Murante G.,   Springel V.,  2009, \mn@doi [Monthly
  Notices of the Royal Astronomical Society]
  {10.1111/j.1365-2966.2009.15034.x}, 399, 497

\bibitem[\protect\citeauthoryear{Durier \& Dalla~Vecchia}{Durier \&
  Dalla~Vecchia}{2012}]{Durier2012}
Durier F.,  Dalla~Vecchia C.,  2012, \mn@doi [Monthly Notices of the Royal
  Astronomical Society] {10.1111/j.1365-2966.2011.19712.x}, 419, 465

\bibitem[\protect\citeauthoryear{Evrard, Summers  \& Davis}{Evrard
  et~al.}{1994}]{Evrard1994}
Evrard A.~E.,  Summers F.~J.,   Davis M.,  1994, \mn@doi [The Astrophysical
  Journal] {10.1086/173700}, 422, 11

\bibitem[\protect\citeauthoryear{Gingold \& Monaghan}{Gingold \&
  Monaghan}{1977}]{Gingold1977}
Gingold R.~A.,  Monaghan J.~J.,  1977, \mn@doi [Monthly Notices of the Royal
  Astronomical Society] {10.1093/mnras/181.3.375}, 181, 375

\bibitem[\protect\citeauthoryear{Harris et~al.,}{Harris
  et~al.}{2020}]{Harris2020}
Harris C.~R.,  et~al., 2020, arXiv e-prints, p. arXiv:2006.10256

\bibitem[\protect\citeauthoryear{Hernquist \& Katz}{Hernquist \&
  Katz}{1989}]{Hernquist1989}
Hernquist L.,  Katz N.,  1989, \mn@doi [The Astrophysical Journal Supplement
  Series] {10.1086/191344}, 70, 419

\bibitem[\protect\citeauthoryear{Hopkins}{Hopkins}{2013}]{Hopkins2013}
Hopkins P.~F.,  2013, \mn@doi [Monthly Notices of the Royal Astronomical
  Society] {10.1093/mnras/sts210}, 428, 2840

\bibitem[\protect\citeauthoryear{Hopkins}{Hopkins}{2015}]{Hopkins2015}
Hopkins P.~F.,  2015, \mn@doi [Monthly Notices of the Royal Astronomical
  Society] {10.1093/mnras/stv195}, 450, 53

\bibitem[\protect\citeauthoryear{Hopkins et~al.,}{Hopkins
  et~al.}{2018}]{Hopkins2018}
Hopkins P.~F.,  et~al., 2018, \mn@doi [Monthly Notices of the Royal
  Astronomical Society] {10.1093/mnras/sty1690}, 480, 800

\bibitem[\protect\citeauthoryear{Hu, Naab, Walch, Moster  \& Oser}{Hu
  et~al.}{2014}]{Hu2014}
Hu C.-Y.,  Naab T.,  Walch S.,  Moster B.~P.,   Oser L.,  2014, \mn@doi
  [Monthly Notices of the Royal Astronomical Society] {10.1093/mnras/stu1187},
  443, 1173

\bibitem[\protect\citeauthoryear{Hunter}{Hunter}{2007}]{Hunter2007}
Hunter J.~D.,  2007, \mn@doi [Computing in Science \& Engineering]
  {10.1109/MCSE.2007.55}, 9, 90

\bibitem[\protect\citeauthoryear{Lam, Pitrou  \& Seibert}{Lam
  et~al.}{2015}]{Lam2015}
Lam S.~K.,  Pitrou A.,   Seibert S.,  2015, in Proceedings of the Second
  Workshop on the {{LLVM}} Compiler Infrastructure in {{HPC}}. {{LLVM}} '15.
{Association for Computing Machinery}, {New York, NY, USA},
  \mn@doi{10.1145/2833157.2833162}

\bibitem[\protect\citeauthoryear{Marinacci, Sales, Vogelsberger, Torrey  \&
  Springel}{Marinacci et~al.}{2019}]{Marinacci2019}
Marinacci F.,  Sales L.~V.,  Vogelsberger M.,  Torrey P.,   Springel V.,  2019,
  arXiv:1905.08806 [astro-ph]

\bibitem[\protect\citeauthoryear{McCarthy, Schaye, Bird  \& Le~Brun}{McCarthy
  et~al.}{2017}]{McCarthy2017}
McCarthy I.~G.,  Schaye J.,  Bird S.,   Le~Brun A. M.~C.,  2017, \mn@doi
  [Monthly Notices of the Royal Astronomical Society] {10.1093/mnras/stw2792},
  465, 2936

\bibitem[\protect\citeauthoryear{Monaghan}{Monaghan}{1992}]{Monaghan1992}
Monaghan J.~J.,  1992, \mn@doi [Annual Review of Astronomy and Astrophysics]
  {10.1146/annurev.aa.30.090192.002551}, 30, 543

\bibitem[\protect\citeauthoryear{Monaghan \& Gingold}{Monaghan \&
  Gingold}{1983}]{Monaghan1983}
Monaghan J.,  Gingold R.,  1983, \mn@doi [Journal of Computational Physics]
  {10.1016/0021-9991(83)90036-0}, 52, 374

\bibitem[\protect\citeauthoryear{Morris \& Monaghan}{Morris \&
  Monaghan}{1997}]{Morris1997}
Morris J.,  Monaghan J.,  1997, \mn@doi [Journal of Computational Physics]
  {10.1006/jcph.1997.5690}, 136, 41

\bibitem[\protect\citeauthoryear{Navarro \& White}{Navarro \&
  White}{1993}]{Navarro1993}
Navarro J.~F.,  White S. D.~M.,  1993, \mn@doi [\textbackslash mnras]
  {10.1093/mnras/265.2.271}, 265, 271

\bibitem[\protect\citeauthoryear{Price}{Price}{2008}]{Price2008}
Price D.~J.,  2008, \mn@doi [Journal of Computational Physics]
  {10.1016/j.jcp.2008.08.011}, 227, 10040

\bibitem[\protect\citeauthoryear{Price}{Price}{2012}]{Price2012}
Price D.~J.,  2012, \mn@doi [Journal of Computational Physics]
  {10.1016/j.jcp.2010.12.011}, 231, 759

\bibitem[\protect\citeauthoryear{Read \& Hayfield}{Read \&
  Hayfield}{2012}]{Read2012}
Read J.~I.,  Hayfield T.,  2012, \mn@doi [Monthly Notices of the Royal
  Astronomical Society] {10.1111/j.1365-2966.2012.20819.x}, 422, 3037

\bibitem[\protect\citeauthoryear{Ritchie \& Thomas}{Ritchie \&
  Thomas}{2001}]{Ritchie2001}
Ritchie B.~W.,  Thomas P.~A.,  2001, \mn@doi [Monthly Notices of the Royal
  Astronomical Society] {10.1046/j.1365-8711.2001.04268.x}, 323, 743

\bibitem[\protect\citeauthoryear{Rosswog}{Rosswog}{2019}]{Rosswog2019}
Rosswog S.,  2019, arXiv:1911.13093 [astro-ph, physics:physics]

\bibitem[\protect\citeauthoryear{Saitoh \& Makino}{Saitoh \&
  Makino}{2013}]{Saitoh2013}
Saitoh T.~R.,  Makino J.,  2013, \mn@doi [The Astrophysical Journal]
  {10.1088/0004-637X/768/1/44}, 768, 44

\bibitem[\protect\citeauthoryear{Schaller, Dalla~Vecchia, Schaye, Bower,
  Theuns, Crain, Furlong  \& McCarthy}{Schaller et~al.}{2015}]{Schaller2015}
Schaller M.,  Dalla~Vecchia C.,  Schaye J.,  Bower R.~G.,  Theuns T.,  Crain
  R.~A.,  Furlong M.,   McCarthy I.~G.,  2015, \mn@doi [Monthly Notices of the
  Royal Astronomical Society] {10.1093/mnras/stv2169}, 454, 2277

\bibitem[\protect\citeauthoryear{Schaye et~al.,}{Schaye
  et~al.}{2015}]{Schaye2015}
Schaye J.,  et~al., 2015, \mn@doi [Monthly Notices of the Royal Astronomical
  Society] {10.1093/mnras/stu2058}, 446, 521

\bibitem[\protect\citeauthoryear{{SciPy 1.0 Contributors} et~al.,}{{SciPy 1.0
  Contributors} et~al.}{2020}]{SciPy1.0Contributors2020}
{SciPy 1.0 Contributors} et~al., 2020, \mn@doi [Nature Methods]
  {10.1038/s41592-019-0686-2}, 17, 261

\bibitem[\protect\citeauthoryear{Springel}{Springel}{2005}]{Springel2005}
Springel V.,  2005, \mn@doi [Monthly Notices of the Royal Astronomical Society]
  {10.1111/j.1365-2966.2005.09655.x}, 364, 1105

\bibitem[\protect\citeauthoryear{Springel \& Hernquist}{Springel \&
  Hernquist}{2002}]{Springel2002}
Springel V.,  Hernquist L.,  2002, \mn@doi [Monthly Notices of the Royal
  Astronomical Society] {10.1046/j.1365-8711.2002.05445.x}, 333, 649

\bibitem[\protect\citeauthoryear{Steinwandel, Moster, Naab, Hu  \&
  Walch}{Steinwandel et~al.}{2020}]{Steinwandel2020}
Steinwandel U.~P.,  Moster B.~P.,  Naab T.,  Hu C.-Y.,   Walch S.,  2020,
  \mn@doi [Monthly Notices of the Royal Astronomical Society]
  {10.1093/mnras/staa821}, 495, 1035

\bibitem[\protect\citeauthoryear{Stern, Fielding, {Faucher-Gigu{\`e}re}  \&
  Quataert}{Stern et~al.}{2019}]{Stern2019}
Stern J.,  Fielding D.,  {Faucher-Gigu{\`e}re} C.-A.,   Quataert E.,  2019,
  arXiv:1906.07737 [astro-ph]

\bibitem[\protect\citeauthoryear{Teklu, Remus, Dolag, Beck, Burkert, Schmidt,
  Schulze  \& Steinborn}{Teklu et~al.}{2015}]{Teklu2015}
Teklu A.~F.,  Remus R.-S.,  Dolag K.,  Beck A.~M.,  Burkert A.,  Schmidt A.~S.,
   Schulze F.,   Steinborn L.~K.,  2015, \mn@doi [The Astrophysical Journal]
  {10.1088/0004-637X/812/1/29}, 812, 29

\bibitem[\protect\citeauthoryear{Teyssier}{Teyssier}{2002}]{Teyssier2002}
Teyssier R.,  2002, \mn@doi [Astronomy \& Astrophysics]
  {10.1051/0004-6361:20011817}, 385, 337

\bibitem[\protect\citeauthoryear{Tremmel, Karcher, Governato, Volonteri, Quinn,
  Pontzen, Anderson  \& Bellovary}{Tremmel et~al.}{2017}]{Tremmel2017}
Tremmel M.,  Karcher M.,  Governato F.,  Volonteri M.,  Quinn T.~R.,  Pontzen
  A.,  Anderson L.,   Bellovary J.,  2017, \mn@doi [Monthly Notices of the
  Royal Astronomical Society] {10.1093/mnras/stx1160}, 470, 1121

\bibitem[\protect\citeauthoryear{Tumlinson, Peeples  \& Werk}{Tumlinson
  et~al.}{2017}]{Tumlinson2017}
Tumlinson J.,  Peeples M.~S.,   Werk J.~K.,  2017, \mn@doi [Annual Review of
  Astronomy and Astrophysics] {10.1146/annurev-astro-091916-055240}, 55, 389

\bibitem[\protect\citeauthoryear{Vogelsberger et~al.,}{Vogelsberger
  et~al.}{2014}]{Vogelsberger2014}
Vogelsberger M.,  et~al., 2014, \mn@doi [Monthly Notices of the Royal
  Astronomical Society] {10.1093/mnras/stu1536}, 444, 1518

\bibitem[\protect\citeauthoryear{Vogelsberger, Marinacci, Torrey  \&
  Puchwein}{Vogelsberger et~al.}{2020}]{Vogelsberger2020}
Vogelsberger M.,  Marinacci F.,  Torrey P.,   Puchwein E.,  2020, \mn@doi
  [Nature Reviews Physics] {10.1038/s42254-019-0127-2}, 2, 42

\bibitem[\protect\citeauthoryear{{van Rossum} \& Drake~Jr}{{van Rossum} \&
  Drake~Jr}{1995}]{VanRossum1995}
{van Rossum} G.,  Drake~Jr F.~L.,  1995, Python Tutorial.
 Vol. 620, {Centrum voor Wiskunde en Informatica}, {Amsterdam}

\makeatother
\end{thebibliography}

\end{document}